\documentclass[10pt,journal]{IEEEtran}
\IEEEoverridecommandlockouts
\usepackage{cite}
\usepackage{enumitem}
\usepackage{amsmath,amssymb,amsfonts}
\usepackage{graphicx}
\usepackage{amsmath}
\usepackage{xcolor}
\usepackage{multirow}
\usepackage{amssymb}
\usepackage{amsthm}

\newtheorem*{theorem*}{Theorem}

\newtheorem*{lemma*}{Lemma}
\usepackage{tabularx}
\usepackage{url}

\usepackage{algorithm}
\usepackage{algpseudocode}
\makeatletter
\def\therule{\makebox[\algorithmicindent][l]{\hspace*{.5em}\vrule height .75\baselineskip depth .25\baselineskip}}%

\newtoks\therules
\therules={}
\def\appendto#1#2{\expandafter#1\expandafter{\the#1#2}}
\def\gobblefirst#1{
	#1\expandafter\expandafter\expandafter{\expandafter\@gobble\the#1}}%
\def\LState{\State\unskip\the\therules}
\def\pushindent{\appendto\therules\therule}%
\def\popindent{\gobblefirst\therules}%
\def\printindent{\unskip\the\therules}%
\def\printandpush{\printindent\pushindent}%
\def\popandprint{\popindent\printindent}%
\algdef{SE}[WHILE]{While}{EndWhile}[1]
{\printandpush\algorithmicwhile\ #1\ \algorithmicdo}
{\popandprint\algorithmicend\ \algorithmicwhile}%
\algdef{SE}[FOR]{For}{EndFor}[1]
{\printandpush\algorithmicfor\ #1\ \algorithmicdo}
{\popandprint\algorithmicend\ \algorithmicfor}%
\algdef{S}[FOR]{ForAll}[1]
{\printindent\algorithmicforall\ #1\ \algorithmicdo}%
\algdef{SE}[LOOP]{Loop}{EndLoop}
{\printandpush\algorithmicloop}
{\popandprint\algorithmicend\ \algorithmicloop}%
\algdef{SE}[REPEAT]{Repeat}{Until}
{\printandpush\algorithmicrepeat}[1]
{\popandprint\algorithmicuntil\ #1}%
\algdef{SE}[IF]{If}{EndIf}[1]
{\printandpush\algorithmicif\ #1\ \algorithmicthen}
{\popandprint\algorithmicend\ \algorithmicif}%
\algdef{C}[IF]{IF}{ElsIf}[1]
{\popandprint\pushindent\algorithmicelse\ \algorithmicif\ #1\ \algorithmicthen}%
\algdef{Ce}[ELSE]{IF}{Else}{EndIf}
{\popandprint\pushindent\algorithmicelse}%
\algdef{SE}[PROCEDURE]{Procedure}{EndProcedure}[2]
{\printandpush\algorithmicprocedure\ \textproc{#1}\ifthenelse{\equal{#2}{}}{}{(#2)}}%
{\popandprint\algorithmicend\ \algorithmicprocedure}%
\algdef{SE}[FUNCTION]{Function}{EndFunction}[2]
{\printandpush\algorithmicfunction\ \textproc{#1}\ifthenelse{\equal{#2}{}}{}{(#2)}}%
{\popandprint\algorithmicend\ \algorithmicfunction}%

\usepackage{array}
\newcolumntype{P}[1]{>{\centering\arraybackslash}p{#1}}

\usepackage{enumitem}
\newlist{legal}{enumerate}{10}
\setlist[legal]{label*=\arabic*.}

\newtheorem*{definition*}{Definition}

\newtheorem*{condition*}{Condition}
\newtheorem*{proof*}{Proof}

\usepackage{mathtools}
\newsavebox\myboxA
\newsavebox\myboxB
\newlength\mylenA

\newcommand{\subparagraph}{}

\newcommand*\xoverline[2][0.75]{%
	\sbox{\myboxA}{$\m@th#2$}%
	\setbox\myboxB\null
	\ht\myboxB=\ht\myboxA%
	\dp\myboxB=\dp\myboxA%
	\wd\myboxB=#1\wd\myboxA
	\sbox\myboxB{$\m@th\overline{\copy\myboxB}$}
	\setlength\mylenA{\the\wd\myboxA}
	\addtolength\mylenA{-\the\wd\myboxB}%
	\ifdim\wd\myboxB<\wd\myboxA%
	\rlap{\hskip 0.5\mylenA\usebox\myboxB}{\usebox\myboxA}%
	\else
	\hskip -0.5\mylenA\rlap{\usebox\myboxA}{\hskip 0.5\mylenA\usebox\myboxB}%
	\fi}
\makeatother

\usepackage{scalerel}[2014/03/10]
\usepackage{stackengine}

\newlength\myindent
\usepackage{epsfig}
\usepackage{tabularx}
\usepackage{supertabular,booktabs}
\usepackage{capt-of}

\usepackage{longtable}
\usepackage{siunitx}
\usepackage{booktabs}

\usepackage{subcaption}

\usepackage{algorithm}
\usepackage{algpseudocode}
\renewcommand{\algorithmicforall}{\textbf{for each}}

\algrenewcommand\algorithmicrequire{\textbf{Precondition:}}
\algrenewcommand\algorithmicensure{\textbf{Postcondition:}}

\usepackage{etoolbox}
\usepackage{textcomp}
\usepackage[normalem]{ulem}
\usepackage{mathtools}
\usepackage{float}
\usepackage{tikz}
\usetikzlibrary{shapes.geometric, arrows}
\tikzstyle{startstop} = [rectangle, rounded corners, minimum width=3cm, minimum height=1cm,text centered, draw=black, fill=red!30]
\tikzstyle{io} = [trapezium, trapezium left angle=70, trapezium right angle=110, minimum width=3cm, minimum height=1cm, text centered, text width=3cm, draw=black, fill=blue!30]
\tikzstyle{process} = [rectangle, minimum width=3cm, minimum height=1cm, text centered, draw=black, fill=orange!30]
\tikzstyle{decision} = [diamond, minimum width=2.5cm, minimum height=2.5cm, text centered, draw=black, fill=green!30]
\tikzstyle{arrow} = [thick,->,>=stealth]
\tikzstyle{process} = [rectangle, minimum width=3cm, minimum height=1cm, text centered, text width=3cm, draw=black, fill=orange!30]
\def\BibTeX{{\rm B\kern-.05em{\sc i\kern-.025em b}\kern-.08em
		T\kern-.1667em\lower.7ex\hbox{E}\kern-.125emX}}
		
\usepackage[compact]{titlesec}
\titlespacing{\section}{0pt}{*0}{*0}
\titlespacing{\subsection}{0pt}{*0}{*0}
\titlespacing{\subsubsection}{0pt}{*0}{*0}
	
\begin{document}
	\bstctlcite{IEEEexample:BSTcontrol}
	
	\title{Curriculum Based Reinforcement Learning of Grid Topology Controllers to Prevent Thermal Cascading
	}

\makeatletter	
\patchcmd{\@maketitle}
  {\addvspace{0.5\baselineskip}\egroup}
  {\addvspace{-1\baselineskip}\egroup}
  {}
  {}

	\author{
		\IEEEauthorblockN{Amarsagar Reddy Ramapuram Matavalam},
		\IEEEauthorblockA{\textit{Member}},
		\textit{IEEE},
		\and
		\IEEEauthorblockN{Kishan Prudhvi Guddanti},
		\IEEEauthorblockA{\textit{Student Member}},
		\textit{IEEE},
		\IEEEauthorblockN{Yang Weng},
		\IEEEauthorblockA{\textit{Senior Member}},
		\textit{IEEE}, and
		\and
		\IEEEauthorblockN{Venkataramana Ajjarapu},
		\IEEEauthorblockA{\textit{Fellow}},
		\textit{IEEE}
	}
	
	
	\maketitle
	\begin{abstract}
	This paper describes how domain knowledge of power system operators can be integrated into reinforcement learning (RL) frameworks to effectively learn agents that control the grid's topology to prevent thermal cascading. Typical RL-based topology controllers fail to perform well due to the large search/optimization space. Here, we propose an actor-critic-based agent to address the problem's combinatorial nature and train the agent using the RL environment developed by RTE, the French TSO. To address the challenge of the large optimization space, a curriculum-based approach with reward tuning is incorporated into the training procedure by modifying the environment using network physics for enhanced agent learning. Further, a parallel training approach on multiple scenarios is employed to avoid biasing the agent to a few scenarios and make it robust to the natural variability in grid operations. Without these modifications to the training procedure, the RL agent failed for most test scenarios, illustrating the importance of properly integrating domain knowledge of physical systems for real-world RL learning. The agent was tested by RTE for the 2019 learning to run the power network challenge and was awarded the $2^{nd}$ place in accuracy and $1^{st}$ place in speed. The developed code is open-sourced for public use.

	\end{abstract}
	
	\section{Introduction}
	\label{sec:intro}
	Grid operators need to ensure that line currents do not exceed physical limits. If left unattended or an appropriate response is delayed, then overloaded lines could lead to cascading due to line thermal limit violations \cite{hines2009cascading}. Transmission system operators (TSOs) prefer an economical and flexible solution like dynamic topology reconfiguration that uses existing infrastructure over the other solutions like load shedding, peak shaving, curtailment, transmission expansion planning \cite{fisher2008optimal, soroush2013accuracies, karangelos2021cooperative}. Even though dynamic topology reconfiguration is preferred by the TSOs \cite{karangelos2021cooperative}, it is still beyond the state-of-the-art to optimally control the grid topology ``at
scale”, beyond the level of ``transmission line switching” operation \cite{fisher2008optimal}. For example, implementation of ``bus splitting/merging" operation (node reconfiguration at a substation using the node-breaker model) ``at scale" is non-trivial due to the nonlinear combinatorial nature of the graph-like structure of the power grids \cite{marot2020learning}.

\cite{marot2018expert} proposed an expert system-based approach that incorporates both transmission line switching and bus splitting/merging operations. This expert system-based approach is sufficiently fast but suffers from accuracy issues at times, and also, it cannot account for the impact of an optimal control action over a time horizon \cite{marot2020learning}. To solve this issue, controllers for dynamic topology reconfiguration are developed to provide optimal control actions over a time horizon. \cite{granelli2006optimal} includes the time horizon concept but uses a mixed-integer nonlinear optimization method which takes longer times to solve. \cite{schnyder1988integrated, schnyder1990security} proposed a fast method for topology reconfiguration, but due to the problem's large search space, they do not look for the optimal control actions.

Recently, with interest to develop real-time recommendation systems, artificial intelligence (AI) based controllers have been of interest to the industry \cite{karangelos2021cooperative, marot2020learning}. AI is used in diverse applications by the industry and few such examples are power grid voltage control \cite{duan2019deep}, stability \cite{you2020review}, emergency load shedding \cite{li2021research}, energy storage systems \cite{wang2021deep}. \cite{subramanian2021exploring} proposes a reinforcement learning-based topology controller but training such a controller to perform well over a wide range of operating scenarios is non-trivial. \cite{marot2020learning} overcomes such issues by training with more scenarios that resemble real-world behaviors. However, the controllers developed in \cite{marot2020learning} cannot account for large grids and do not optimize for the line losses.

In this work, a systematic approach to develop topology controllers that plan over a time horizon is proposed. The advantage of the proposed method is that it uses domain knowledge to make the AI-based controller learn well, even in the case of a large solution search space. We propose an actor-critic (A3C) topology controller that can learn by deploying multiple agents in parallel worlds/environments and aggregating the learned policies into a single agent. Furthermore, to simplify the hard-to-solve learning process of developing topology controllers for power grids, we propose power grid domain-specific curriculum learning strategies that can improve any arbitrary AI-based controller's performance and training time.

The contributions of this work are
\begin{itemize}
\item A physics-based state selection and reward design that can enable the learning of the A3C-based topology controller to prevent thermal cascading. 
\item A curriculum-learning strategy for accelerating A3C-controller learning with the potential to generalize to other sequential network flow planning problems. 
\item Testing and validating the proposed curriculum approach on the IEEE 14-bus system and comparing its behavior with a non-machine learning forecast-based agent and an out-of-the-box RL agent. The proposed method outperforms the other agents because of the domain knowledge embedded in the curriculum strategy. 
\item Open-sourced code that implements the curriculum learning along with the physics-based reward function and state selection. The agent learnt by this code placed $2^{nd}$ in accuracy and $1^{st}$ in speed in the L2RPN-2019 competition \cite{marot2020learning}.

\end{itemize}
The rest of the paper is organized as follows. Section II describes the problem of managing the transmission line congestion and formulates it as a sequential decision-making problem that can be solved by reinforcement learning; Section III describes the general advantage-actor-critic architecture and the training procedure. Section IV describes the challenges in training an A3C grid topology controller and the modified reward to enable agent learning. Section V describes the physics-inspired curriculum-based approach to accelerate the learning of the A3C agent. Section VI presents the simulation results of the trained RL agent using the curriculum approach; Section VII concludes the paper. 
	\section{Problem Description: Managing the Transmission Line Congestion of Power Grids}
	\label{sec:problem_description}
	In this section, first, we introduce the problem of transmission line congestion which causes a cascading event that may result in the blackout of the power grid. Second, to manage such congestion in the power grids, we briefly mention the various preventive techniques and introduce ``actions" (real-time topology switching) that are flexible as well as cost-effective from the power grid operator's perspective \cite{fisher2008optimal}. Third, we formulate this energy management of power network as a dynamic/sequential planning problem using an objective function and set of constraints. Finally, the complexity of the formulated optimization problem and the size of search space is presented as motivation to ``learn" a ``policy" (sequence of actions) for the real-time oriented control solution.
\subsection{Black out of power grids due to cascading events}
\label{subsec:power_grid_intro_example}
\begin{figure}
    \centering
    \centerline{\includegraphics[width=\linewidth]{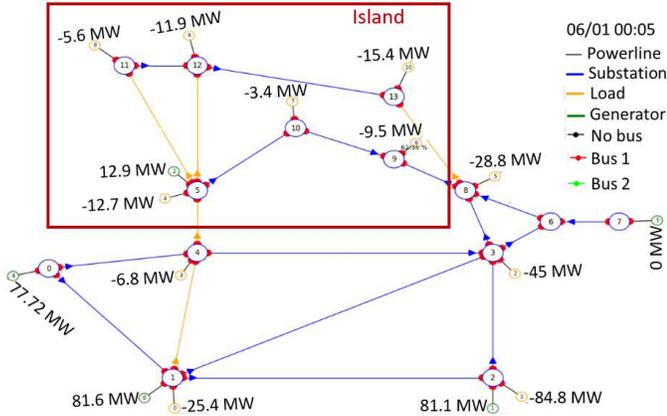}} 
    \caption{Power grid cascade event when ``chronics" are injected to $14$ bus system designed by the L2RPN competition \cite{lerousseau2021design, grid2op}.}
    \label{fig:cascade_pic}
    \vspace{-2em}
\end{figure}

In this subsection, we present the $14$ bus system designed by \cite{lerousseau2021design} to demonstrate the problem of maximizing the transfer capability of the power grid while avoiding the cascading events over a time horizon. Fig.~\ref{fig:cascade_pic} presents a $14$-bus system with $14$ substations, $20$ transmission lines and $16$ injections (both generations and loads combines). In Fig.~\ref{fig:cascade_pic}, the substations are indicated by the nodes (blue circles) in the graph; the yellow circles indicate loads, and the green circles indicate generations. Additionally, as shown in the legend of Fig.~\ref{fig:cascade_pic}, each substation has two bus bars, namely ``bus $1$" and ``bus $2$". An element (either a line or load or generator) can be located at a substation connected to either ``bus $1$" or ``bus $2$" (node breaker model). Furthermore, realistic design is constructed by assigning generators with IDs $0, 1, 2, 3, 4, 5$ with nuclear, thermal, wind, solar, solar, and hydro generation profiles.

To represent the realistic power grid operation scenario, realistic generation and load consumption profiles (\textit{hereafter referred to as ``chronics"}) are injected into the power grid for 2000 time-steps of 5 minutes each (equal to 1 week). \textit{In the interest of space, the generation and load profiles are not shown in this initial draft}. 
The transmission lines in this $14$ bus system have thermal limits, and when a transmission line is overloaded, it will disconnect and become out-of-service. 

Without performing any modification to the given topology presented in Fig.~\ref{fig:cascade_pic}, the injections into the grid result in a cascading event that leads to power grid blackout. One such cascading event is as follows; first, the transmission line connecting substations $1$ and $4$ are overloaded and becomes out-of-service. The loss of this line reduces the power grid's overall transfer capability, which in turn overloads the other transmission lines in the power grid. This overloading causes the disconnection of the transmission line $4-5$ two time steps later. Finally, the transmission lines  $8-9$ and $8-13$ disconnect simultaneously the next time step due to high line loading of $311.72\%$ and $174.11\%$ respectively, resulting in an island as shown in Fig.~\ref{fig:cascade_pic}. However, the formation of islands is not a necessary condition for the blackout of the power grid, and blackout can also occur due to voltage instability condition \cite{lerousseau2021design} which is identified by the lack of a solution for a specific set of injections. Hence, it is equally essential to consider cascades that create islands (network flow problem) and voltage stability conditions to ensure power flow solution exists when managing the power flows in the power grid.

\subsection{Topology switching actions}
\label{subsec:action_space_intro}

The current focus of the industry is to not only manage power flows in the grid to avoid cascading events as described in Section.~\ref{subsec:power_grid_intro_example}, but it is also to maximize the transfer capability of the power grid by minimizing the line losses \cite{marot2020learning}. Both industry and academia have provided many preventive actions based solutions for congestion management and loss minimization problems. Some of them are new transmission lines (transmission expansion problem), reactive power support, transmission switching, etc. However, installation of new equipment on the power grid is not only expensive, but public acceptance is also a growing concern \cite{marot2020learning}. Thus, it is preferred to optimize the operation using the flexibility of existing infrastructure. One such method that is both cost-efficient and flexible is the dynamic reconfiguration of grid topology. 

\begin{figure}
		\centering
        \begin{subfigure}{0.15\textwidth}
    		\includegraphics[width=\linewidth]{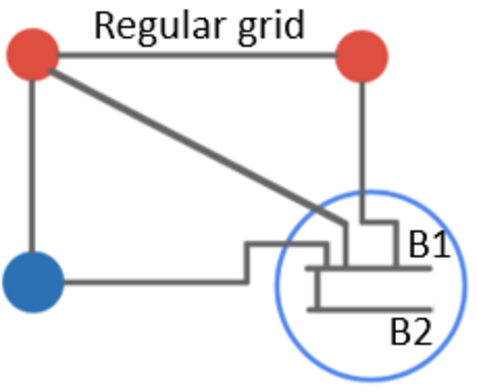}
    		\caption{Power grid with nominal bus bar connectivity.}
    		\label{fig:bus_split_1}
		\end{subfigure}
        ~
		\begin{subfigure}{0.15\textwidth}
			\includegraphics[width=\linewidth]{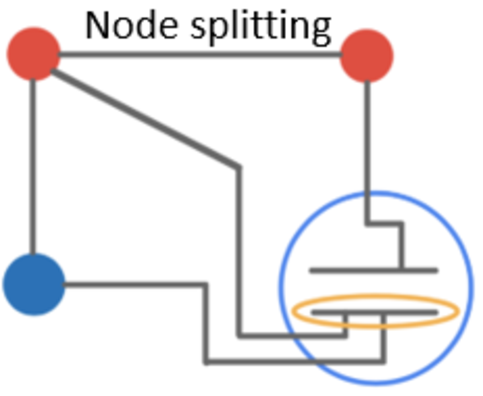}
			\caption{Bus splitting action using the bus bar connectivity.}
			\label{fig:bus_split_2}
		\end{subfigure}
		~
		\begin{subfigure}{0.15\textwidth}			\includegraphics[width=\linewidth]{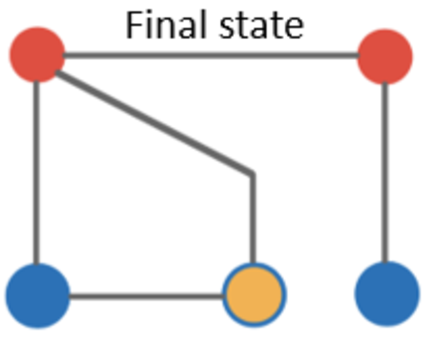}
			\vspace{-2mm}
			\caption{Final state of the topology after a bus splitting action.}
			\label{fig:bus_split_3}
		\end{subfigure}
		\caption{Bus splitting using bus bars in a substation \cite{marot2020learning}.}
		\label{fig:bus_split_full}
		\vspace{-2em}
 \end{figure}

The actions required to implement dynamic reconfiguration of grid topology are 1) transmission line switching and 2) bus splitting/merging using the bus bars in a substation. The transmission line switching action involves the decision to make a line in-service or out-of-service. However, bus splitting actions are more complex, and it is explained using a simple $4$-bus system from Fig.~\ref{fig:bus_split_full}. Fig.~\ref{fig:bus_split_1} presents $4$-bus system with five transmission lines and four substations. Each substation in the network has two bus bars to which the power network elements such as loads, generators, transformers, shunt admittances, and transmission lines are connected. Fig.~\ref{fig:bus_split_1} shows a topology with three transmission lines connected to the bus bar $1$ (B1) and $2$ (B2). For example, as shown in Fig.~\ref{fig:bus_split_2}, a bus splitting action can be triggered to connect two incoming transmission lines to bus bar $2$ (B2) and one transmission line to bus bar $1$ (B1) separately. This results in a new topology with five nodes, as shown in Fig.~\ref{fig:bus_split_3}, and the new topology can have very different power flow routing properties compared to the original topology. 

\subsection{Model formulation: topology controllers for power grids}
\label{subsec:optimization_formulation}
This subsection discusses the topology controller problem formulation from a traditional optimization approach as a large-scale mixed-integer non-linear programming problem. The difficulty in solving this problem motivates the need for state-of-the-art dynamic optimization techniques. Finally, we provide a concise mathematical representation of the topology controllers for power grids solved in this paper.

The objective of a topology controller for the power grid involves identifying the optimal topology grid configuration (combinatorial) that minimizes the total line loading on the power grid to avoid the formation of islands. This objective must be achieved while ensuring that a power flow solution exists for the optimized grid topology with line currents below thermal limits.

It is shown in \cite{pourahmadi2019dynamic} that the traditional optimization formulation of identifying optimal topology \textit{\textbf{at a given snapshot}} is a large scale non-convex mixed-integer non-linear programming problem (in the interest of space, we did not provide full optimization formulation). This is a computationally intensive optimization problem to solve even for commercial solvers. However, the real-world problem is not a single snapshot problem but rather the optimal topology must be designed considering the variation of load and generator injections \textbf{over a time horizon} (several time steps). This significantly increases the computational complexity of the problem. However, there is a need for real-time/fast optimal topology recommendation systems. 
To address this need, the topology controller problem for power grids is first formulated as a sequential decision-making problem, and then RL agents are trained to solve the problem in real-time using historical data. 
The sequential planning problem is shown below:
\begin{align}
    \mathop {\min }\limits_{{\tau}}  & \;\sum\limits_{t = 1}^{t = n} {\sum\limits_{\forall p \in E} {\left( {\frac{{{I_{p,t}}}}{{{I_{p,max}}}}} \right)} }, \label{obj_fine}\\
    sub.\;to:\;&{f_{{\tau}}}\left( {{x_t}} \right) = 0;\;\forall t = \{ 1,2, \cdots ,n\},\label{acpf}\\
    & T(x_t) \in A(T(x_{t-1}));\forall t = \{ 2, \cdots ,n\},\label{topo_constraint}\\
    &{I_{p,t}} \le {I_{p,max}}\;\forall p \in E;\;\forall t = \{ 1, \cdots ,n\}.\label{current_limits}
\end{align}
The aim of the topology controller is to minimize the total line loading on the grid over a time horizon $t=\{1,2,…,n\}$ (equation \eqref{obj_fine}) by identifying the optimal topology $\tau$ for every time step $t$ with transmission line switching and bus splitting/merging actions. For entire time horizon $t$, $f_{\tau}(x_t )=0$ represents the AC power flow constraint of the power grid with different topologies $\tau$ and state vectors $x_t$. The state vector $x_t$ includes the bus voltages, load injections and generator injections and the grid topology representation. The constraint \eqref{topo_constraint} represents the constraint between topologies in consecutive time steps. The grid topology at a time $t$ $\left(T(x_t)\right)$ should lie in the allowable set of topologies based on the topology at the previous time step $\left( A(T(x_{t-1}))\right)$. The constraint \eqref{current_limits} represents limit on the current magnitude in a transmission line $p$ ($I_{p,t}$). The current must be less than its thermal limits $I_{p,max}$ over the entire time horizon $t$ where $\forall p \in E$ where $E$ is the set of all transmission lines in the power grid.
\subsection{Real-time topology controllers for power grids: size of topology space}
\label{subsec:actions_space_complexity_and_motivation}
The total possible line switching topologies for a grid in Fig.~\ref{fig:cascade_pic} with 20 transmission line is $2^{20}$. Similarly, the total bus splitting/merging topologies at a substation with $k$ elements is $\approx 2^{k-1}$ which equal to $1,397,519,564$ unique topologies for the system in Fig.~\ref{fig:cascade_pic}. Thus, the total number of possible topology configurations available at a given time step is $1,397,519,564 \times 2^{20}$. Most of these topologies are not viable as they lead to islanding or power flow divergence. 
\textit{Thus, the complexity of selecting an optimal topology at a given time step is not trivial, let alone computing the strategy over a time horizon}. Hence, there is a need to ``learn the strategy" to pick optimal topology (considering the several time steps) rather than exhaustive optimization search methodologies whenever a new grid operating conditions is considered. 
The field of reinforcement learning deals with learning controllers (also referred to as agents) for sequential decision processes to achieve a specific objective using techniques from machine learning. The approach to developing such an agent for non-linear sequential planning problems like controlling grid topologies with discrete actions is described in the next section.   
	\section{Deep Reinforcement Learning}
	\label{sec:A3C_Agent}

This section considers a standard reinforcement learning setup where an agent interacts with a power grid environment $\xi$ over a discrete number of time steps. At each time step $t$, let the state of the environment be $s_t$. The agent selects an action $a_t$ from an action set $\textbf{A}$ which is implemented in the environment $\xi$. The environment $\xi$ returns the resulting next state $s_{t+1}$ due to action $a_t$ and a reward $r_{t+1}$. The higher the reward, the better the action $a_t$ corresponds to the state $s_t$. This procedure is repeated until the environment reaches a terminal state.

\begin{figure*}
    \centering
    \centerline{\includegraphics[height = 4cm]{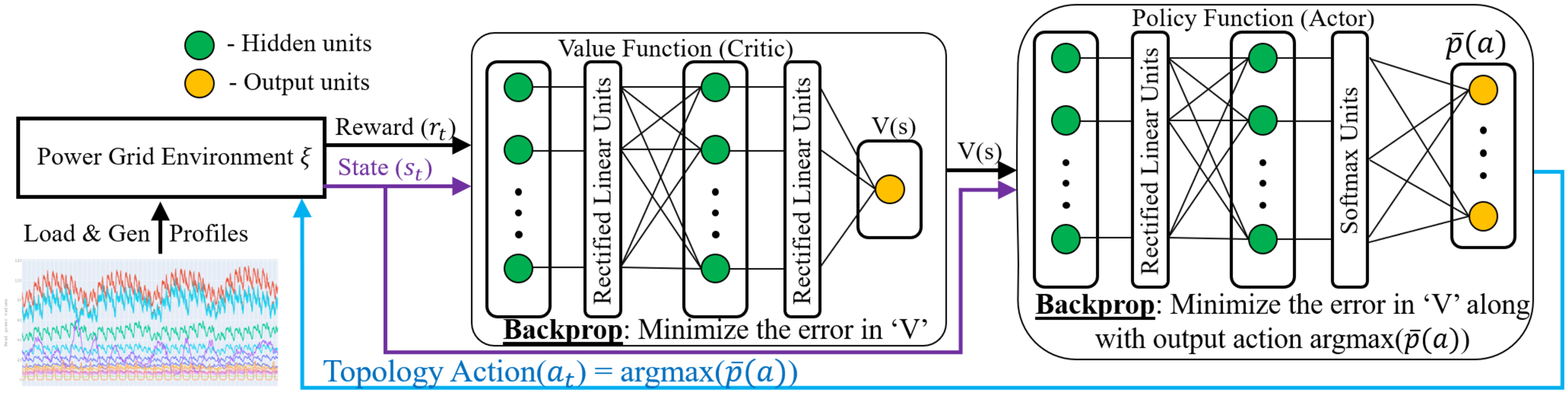}} 
    \caption{Neural network architecture for actor critic based RL agent with continuous state and discrete action spaces.}
    \label{fig:rl_nn_architecture}    \vspace{-0.25in}

\end{figure*}

The proposed deep reinforcement learning agent uses cooperative actor \& critic agents. The actor \& critic are represented as deep neural networks with parameters $\theta$, as shown in Fig.~\ref{fig:rl_nn_architecture}. The actor-critic architecture is valid for discrete action spaces and is appropriate for node-splitting. The size of the output is equal to the number of discrete actions in the system. Given an action $a_{t-1}$ on an environment $\xi$, the critic looks at the next state $s_{t}$ and reward $r_t$ corresponding to action $a_{t-1}$, it then predicts the value $V(s_t)$ for the state $s_t$ (\textit{policy evaluation}). The actor-network then uses the value $V(s_t)$ and state $s_{t}$ as inputs into its neural network, and by using the property of the softmax layer, it outputs the probabilities of each action as $\Bar{p}(a)$. The action with the largest probability, which is equal to $argmax(\Bar{p}(a))$, is selected as the action $a_{t}$ at time $t$. This specific action is then implemented in $\xi$ resulting in the next state and reward. During training, the actor uses the feedback of the critic network to update its weights to output higher probabilities for better actions at a given state.

\subsection{Training the actor-critic agent}
The objective of an RL agent is to maximize the expected reward overall trajectories $\tau$ such that the policy parameter $\theta$ optimizes the total reward from the environment. A trajectory is also known as an episode/scenario which constitutes complete gameplay, i.e., a sequence of actions ($policy$) from the initial state to the terminal state. This is given by
\begin{align}
    \theta^* &= \operatorname*{argmax}_\theta \operatorname*{E}_{\tau \sim \pi_\theta (\tau)} \left[\sum_{t \sim \tau} r(s_t,a_t)\right], \notag \\
    \theta^* &= \operatorname*{argmax}_\theta \operatorname*{E}_{\tau \sim \pi_\theta (\tau)} \left[r(\tau)\right], \notag \\
    \theta^* &= \operatorname*{argmax}_\theta J(\theta). \label{obj_f}
\end{align}

The update of policy parameter during the training process at iteration $k+1$ is given by $\theta^{k+1} = \theta^k + \eta \cdot \nabla_{\theta} J(\theta)$, where $\eta$ is the learning rate. The efficient learning behavior of the actor-critic network involves the better design of the gradient update of the objective function $\nabla_{\theta}J(\theta)$.


\textbf{Policy gradient on objective function:} Usually, \eqref{obj_f} is solved using gradient descent if the desired objective function $J(\theta)$ is represented as an explicit function. However, in reinforcement learning, the objective function includes the dynamics of the environment, which is a black box. To overcome this drawback, we present the standard REINFORCE update for the policy gradient \cite{sutton2000policy}. \cite{sutton2000policy} shows the derivation of \eqref{PG_eq} from \eqref{obj_f}.
\begin{align}
    \nabla_\theta J(\theta) &= \operatorname*{E}_{\tau \sim \pi_\theta (\tau)} \left[r(\tau) \cdot \nabla_\theta log\left(\pi_\theta (\tau)\right) \right]. \label{PG_eq}
\end{align}
\eqref{PG_eq} provides the vanilla gradient update equation for a policy gradient neural network-based RL agent. However, this formulation does not provide an efficient learning/optimization algorithm. Specifically, we include a few modifications to \eqref{PG_eq} in order to make it more efficient by reducing the variance in $\theta$, 
and discounting future rewards.

\textbf{Reducing the variance of network weights using the \textit{advantage}:} The gradient update using \eqref{PG_eq} suffers from high variance, which often results in convergence or bad learning of the reinforcement learning agent. It is not reliable or efficient to reduce the high variance by increasing the batch size. To address this, we subtract a constant baseline value $V^{\pi}(s)$ that is independent of network parameter $\theta$ \cite{mnih2016asynchronous} as shown below. This term $\left( r(\tau)-V^{\pi}(s)\right)$ is known as $advantage$ value and quantifies the improvement in the total reward for implementing a selected action versus taking no action from the policy network $\pi_{\theta}(\tau)$ in a given trajectory $\tau$.
\begin{align}
    \nabla_\theta J(\theta) &= \operatorname*{E}_{\tau \sim \pi_\theta (\tau)} \left[\nabla_\theta log\left(\pi_\theta (\tau)\right)\cdot \left( r(\tau)-V^{\pi}(s)\right)\right], \label{advtg_eq}
\end{align}
where $r(\tau) \neq \sum_{t \sim \tau} r(s_t,a_t)$ but rather the total accumulated reward in a given trajectory discounted such that the actions preformed far away from the current state $s_t$ has minor impact on the reward $r_t$ and similarly the actions implemented in neighborhood of time step $t$ i.e., $\{t-3,t-2,t-1\}$ have a non-zero impact on the reward $r_t$. This is known as time discounting of the rewards. By simulating $N$ trajectories with fixed neural network weights to approximate the $\operatorname*{E}_{\tau \sim \pi_\theta (\tau)}$, the final gradient is given by (\ref{a3c_final}) below where $Q^{\pi}(s,a)=\sum_{t^{'}=t}^{t_{max}} \gamma^{t^{'}-t}\cdot r(a_{i,t^{'}},s_{i,t^{'}})$. $\gamma$ is the discount factor. $t_{max,i}$ is the total time for each trajectory and is the time taken to reach the terminal state. 
\begin{multline}
    \nabla_\theta J(\theta) \approx \dfrac{1}{N}\sum_{i=1}^{N} \sum_{t=1}^{t_{max}} \nabla_\theta log\left(\pi_\theta (a_{i,t}|s_{i,t})\right)\cdot \\ \left(Q^{\pi}(s,a)-V^{\pi}(s)\right). \label{a3c_final}
\end{multline}

The trajectories $\tau$ can be simulated in parallel to exploit the multi-core nature of modern high-performance computing hardware. As each of the trajectories is independently simulated, the neural network weights are updated asynchronously. This training approach combined with the actor-critic with regularized advantages leads to the state-of-the-art asynchronous-advantage-actor-critic (A3C) model. Application of the A3C method to learn grid topology controllers is described in the next section.


	\section{Training A3C Grid Topology Controllers}
	\label{sec:A3C_pypownet_Agent} 
	
The python package PyPOWNET \cite{lerousseau2021design}, developed by the French system operator RTE, is used as the power grid environment to simulate the action of a topology change. The environment uses varying load injection and generation dispatch and the topology resulting from the actions at a time step to estimate the resulting system states such as line flows, number of consecutive time steps in an overload condition, and node voltages. More information on the pypownet package is found in \cite{lerousseau2021design}. Each episode consists of load and generation profiles lasting for one unique week from the year 2016. 

The default reward from the environment is simple - if the action in the previous time step leads to an unexpected episode termination, then the reward is equal to -1. Otherwise, the return is equal to +1.  Unexpected termination occurs when a load/generator is islanded or if the power flow diverges. The islanding can occur due to a bad action in a previous time step or due to line disconnection caused by overloading or by a combination of the two factors. A line disconnection occurs if the actual current exceeds the rating for three consecutive time steps or if the actual current exceeds 1.5 times the rating for a single time step. Thus, the maximum total award occurs when the agent can take actions that make an episode successful for the maximum number of time steps.

We initially trained the A3C agent (Fig.\ref{fig:rl_nn_architecture}) on the IEEE 14-bus system using the binary +1/-1 reward using the full state vector as the input to the RL agent. The total number of node splitting actions equals 312, and the state vector size is equal to 438. Any grid topology can be created by a sequence of the node splitting actions, and so the number of actions is significantly less compared to the number of possible topologies mentioned in Section~\ref{subsec:actions_space_complexity_and_motivation}. This is how 4 bits can be used to represent 16 numbers. 
\textbf{We observed that the agent could not continuously operate the grid for more than 50 time-steps even after training for 5,000 episodes. On closer examination of the actions taken by the agent, we realized that the agent was unable to learn effectively} due to the following reasons
\begin{itemize}
    \item Redundant actions: The equivalence among various node-splitting actions introduces more parameters in the actor neural network slowing its training.
    \item Correlated and unnormalized states: Correlations are present among the states and their values lie in a wide region as they are not normalized. These attributes lead to ill-conditioned gradients and impede agent learning. 
    \item Unsuitable reward: The binary +1/-1 reward signal is not informative enough for the A3C learning as the agent cannot recognize that the line flows should be maintained below a threshold. 
    
\end{itemize}
Thus, RL agents cannot learn to control the grid topology unless these shortcomings are addressed. 
In order to overcome the challenge of training the A3C agent, we (i) simplified the action space by logically analyzing the network structure, (ii) reduced the state space by analyzing correlations among the various states, and identified the appropriately scaled sub-states to be used, (iii) designed a physically meaningful reward function that is an explicit function of the line current flows. These are detailed further in the next subsections.

\subsection{Action Space Reduction}
Due to the symmetry inherent in a substation with two bus bars, a node-splitting action, and its complement will lead to the same line flows. The elements connected to bus 1 in one action are connected to bus 2 in the other action and vice versa. As the power flow equations are invariant to bus numbering permutations, the resulting line flows under these complementary actions are the same. As these two actions are identical, the total number of effective actions in each substation can be halved, \textbf{reducing the total number of node splitting actions from 312 to 156}. 

\subsection{State Space Reduction and Scaling}
The state vector of the environment has detailed information about the state of the grid and includes the following information at each time step - (i) Topology connectivity information such as busbar and substation of each load/gen./line (ii) Power flow information such as load/gen. Injections, line currents, and load/gen. voltages (iii) Forecasted loads' powers, generators' powers \& voltages. To improve the agent learning, a subset of the states should be chosen that retain the key information about the system while also limiting the correlations. The correlation among some states  (e.g., line currents) to other states (e.g., node voltages) is identified using the physics of the system. Other correlations between the states are identified in a data-driven manner by analyzing a few episodes. In the interest of space, only a few correlations are listed below:
\begin{itemize}
    \item Forecasted generator powers(voltages) are correlated with the current generator powers(voltages)
    \item Active(reactive) power flowing into a line correlated with the active(reactive) power flowing out of the line
\end{itemize}
The final states are chosen so that all correlations among the states are minimized and can uniquely represent the grid's state. These final chosen states are individually scaled so that the maximum values of all the states are in the same order of magnitude, thereby enhancing the training of the RL agents. The final states that are used for the training of the RL agents are listed below. \textbf{The final size of the state space is reduced from 438 to 164 - a reduction of more than $\mathbf{2.5x}$}.

\begin{itemize}
    \item The busbar to which each element is connected. 
    \item Line status, line thermal limits and line flows. 
    \item Generator active power and voltage dispatch.
    \item Load active and reactive power demand.
    \item Time-steps before a substation can be controlled. 
\end{itemize}
\textbf{The combined effect of reducing the actions and states significantly cuts the number of NN parameters $\theta$ from $\sim 110k$ to $\sim 50k$. }  

\subsection{Designing a Suitable Reward}
As the RL agents are trained to maximize the total reward over a time horizon, the reward needs to be appropriately defined to reflect the user's intention for the agent. As the grid controllers should maximize the duration of grid operation, the binary +1/-1 reward seems to be sufficient as the maximum cumulative reward ($J(\theta)$) occurs when the agent avoids termination and increases the number of successful steps. However, the discrete reward leads to a difficult optimization problem as the gradient of $J(\theta)$ can be zero over large regions. Furthermore, the reward at a particular step is independent of the state and does not provide any information about the risk of cascading.


As the underlying mechanism for cascading and thermal disconnections is the line current exceeding its limit, a reward function that explicitly uses the line currents and limits is preferred. Further, this reward function should be designed to discourage line overloading and scenario termination. Thus, we define a new reward function, $r(s_t)$ , shown in (\ref{reward_eq}), with these properties. The reward is essentially the sum of line margins for all the lines if no unexpected termination occurs and is a large negative value (-100) if the termination occurs due to islanding or divergence. 

\begin{figure}
    \centering
    \centerline{\includegraphics[width=\linewidth,trim={0.0cm 0.8cm 2.0cm 1.0cm},clip]{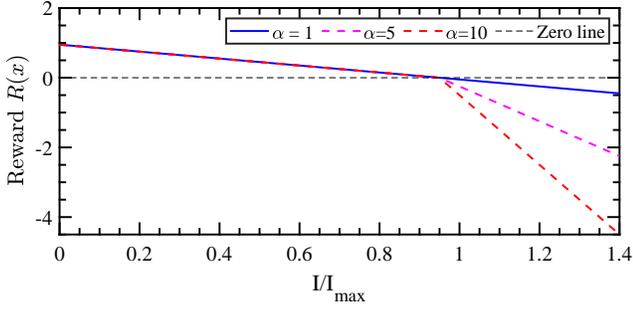}} 
    \caption{{Plot of the function $R(x)$ versus the line current for various parameters of $\alpha$.}}
    \label{fig:reward_func}
    \vspace{-0.25in}
\end{figure}

The function $R(x)$ (shown in Fig.\ref{fig:reward_func}) is a proxy for the line margin and its value reduces as the line loading increases. It is negative if the line current is greater than $0.95$ times its maximum current limit. A threshold of $0.95$ is used instead of $1.0$ during training to 'robustify' the A3C agent and aids in generalizing the A3C agent to similar but unseen states. The value of $\alpha$ in (\ref{reward_eq}) determines the penalty for a line overload. Fig.\ref{fig:reward_func} plots the function $R(x)$ for varying values of $\alpha$. The explicit dependence of the reward on the state facilitates the value function ($V^{\pi}(s)$) learning which in turn makes the learning of the policy $\pi$ easier. This reward is maximum when the A3C agent has the maximum number of continuous successful time steps for all training scenarios with the least line usage, leading to the same outcome of the binary reward. Thus, the solution of (\ref{obj_f}) using the modified reward is also a solution to the (\ref{obj_f}) using the binary reward.

\vspace{-0.2in}
\begin{align}
    r(s_t) &= \left\{ \begin{matrix}
   \sum\limits_{\forall p \in E}{R\left(\frac{{I}_{p,t}}{{I}_{p,max}}\right)}; & $if not terminal time step$  \\
   -100; & $if terminal time step$  \\
\end{matrix} \right.
\label{reward_eq}
\end{align}
\vspace{-0.2in}
\begin{align}
    R(x) &= \left\{ \begin{matrix}
   (0.95-x); & x \leq 0.95  \\
   \alpha \cdot (0.95-x), \alpha \ge 1; & x>0.95  \\
\end{matrix} \right.
\end{align}

\textbf{The identification of bottlenecks impacting the A3C agent learning and developing mitigation strategies resolving them by exploiting the physical understanding of the grid is the first main contribution of the paper.} \textit{This behavior of the reward function was conveyed to RTE and disseminated to a wide audience. As a result, multiple participants of the L2RPN-2020 challenge used this reward function, including the winners.} Implementing the above strategies enables the A3C agent to operate the 14-bus system for more than 200-time steps for a few scenarios after training for 5,000 episodes. However, the learning was very slow, and most of the scenarios failed after 100-time steps. Based on the states and actions analysis, we observed that the agent needs many episodes to learn to avoid risky actions. This is because the number of time steps the environment can operate in a risky topology is typically very low ($<$ 5 time steps) before causing line overloads leading to episode termination. Thus, standard RL training approaches are not designed for learning topology controllers effectively. To address this drawback and accelerate the learning, we developed an efficient training method using physics-inspired curriculum learning to obtain high accuracy for the RL-based topology controllers. This is explained in the next section.

	\section{Physics Inspired Curriculum Strategy for Accelerated Learning}
	\label{sec:curr_training_strategy}
	Curriculum learning is the idea that neural networks learn a difficult task most effectively when first trained on a simpler task.  Curriculum learning is inspired by how humans learn - initially learning simple concepts before attempting complex tasks. It is a form of transfer learning as solving simple tasks is leveraged to solve the more complicated task. A proper curriculum (sequence of tasks with increasing hardness) should be designed to apply this approach for effectively learning grid controllers. Designing an effective curriculum is not easy, and a bad curriculum can impede agent learning. Recent approaches \cite{bengio2009} have proposed to learn curriculum strategies as a part of the overall ML-based approach for classification or regression tasks. In recent years, curriculum learning to accelerate training of RL agents has been explored in various settings \cite{Wu2017Doom}. However, it has not been explored for controlling network flows. \textbf{In this section, we present the physics-inspired curriculum using the behavior of network flows and cascading. The designed curriculum accelerates the A3C agent learning and is the second contribution of the paper. As far as the authors are aware, this is the first time a curriculum has been designed to train RL agents to control network flows.} 

There are a few settings in PyPOWNET that indirectly can increase or reduce the hardness of the environment, as seen by the agent. These configuration parameters are:

\begin{itemize}
    \item The soft overload threshold (SOT), which is the fraction $\frac{{I}_{p,t}}{{I}_{p,max}}$ beyond which an overload alarm is triggered.
    \item The consecutive overload limit (COL) which determines how long a line can be continuously in soft overload before the line is disconnected.
    \item The hard overload threshold (HOT) which is the fraction $\frac{{I}_{p,t}}{{I}_{p,max}}$ beyond which line $p$ immediately disconnects.
\end{itemize}

The default parameters of the environment are SOT = 1.0, COL = 3 time steps and HOT = 1.5. These parameters imply that the overload counter is triggered when the line current exceeds its rating, and the line will be disconnected if the current remains continuously above the \textcolor{black}{SOT} limit for 3 steps (COL). If the line current exceeds 1.5 times the rating (HOT), then it is immediately disconnected. Cascading line outages are the main reason for unexpected termination due to the RL agents, and so initially, we need to prevent cascading in the environment. \textbf{As the problem of cascading occurs due to a sequence of lines disconnecting due to overloads, relaxing the line limit enforcement will directly prevent cascades. It is important to emphasize that the line limits ($I_{p,max}$) are not modified in any of the levels, only the enforcement of the limits is relaxed. Thus, the reward will be negative if the line limits are exceeded. This negative reward will discourage the A3C agents from taking actions that cause line overloads even if the line limit is not enforced.} 

The designed curriculum consists of three levels with increasing difficulty. The environment parameters for the three levels are shown in Table \ref{tab:curr_table}. The $\alpha$ parameter used in the reward function is also increased to ensure that the penalty for overload increases at higher curriculum levels. In Level-1, the SOT is very large ($10^9$), which implies no line limit enforcement. In level-2, the line disconnections are enforced with a large COL of 15. The HOT is very large for this level which prevents immediate line disconnection. Level-3 corresponds to the default environment behavior described above. The levels are designed in a sequential manner that gradually increases the 'strictness' of the enforcement. Next, three propositions are discussed that provide the rationale for improved agent learning with the designed curriculum.

\begin{table}[t]
    \caption{Environment parameters for the curriculum levels.}
    \centering
    \begin{tabular}{|c|c|c|c|c|} 
\hline     
 Level & Reward $\alpha$ & SOT & COL & HOT  \\
\hline   
1 & 1 & $10^9$ & $10^9$ & $10^9$  \\
\hline   
2 & 5 & $2$ & $15$ & $10^9$  \\
\hline   
3 & 10 & $1$ & $3$ & $1.5$  \\

\hline   
\hline   
    \end{tabular}
    \label{tab:curr_table}
    \vspace{-0.15in}

\end{table}

\subsection{Proposition 1: More training samples are seen by the agent for lower levels than higher levels}
\begin{figure}[t]
    \centering
    \centerline{\includegraphics[width=\linewidth,trim={0.0cm 0cm 0.0cm 0},clip]{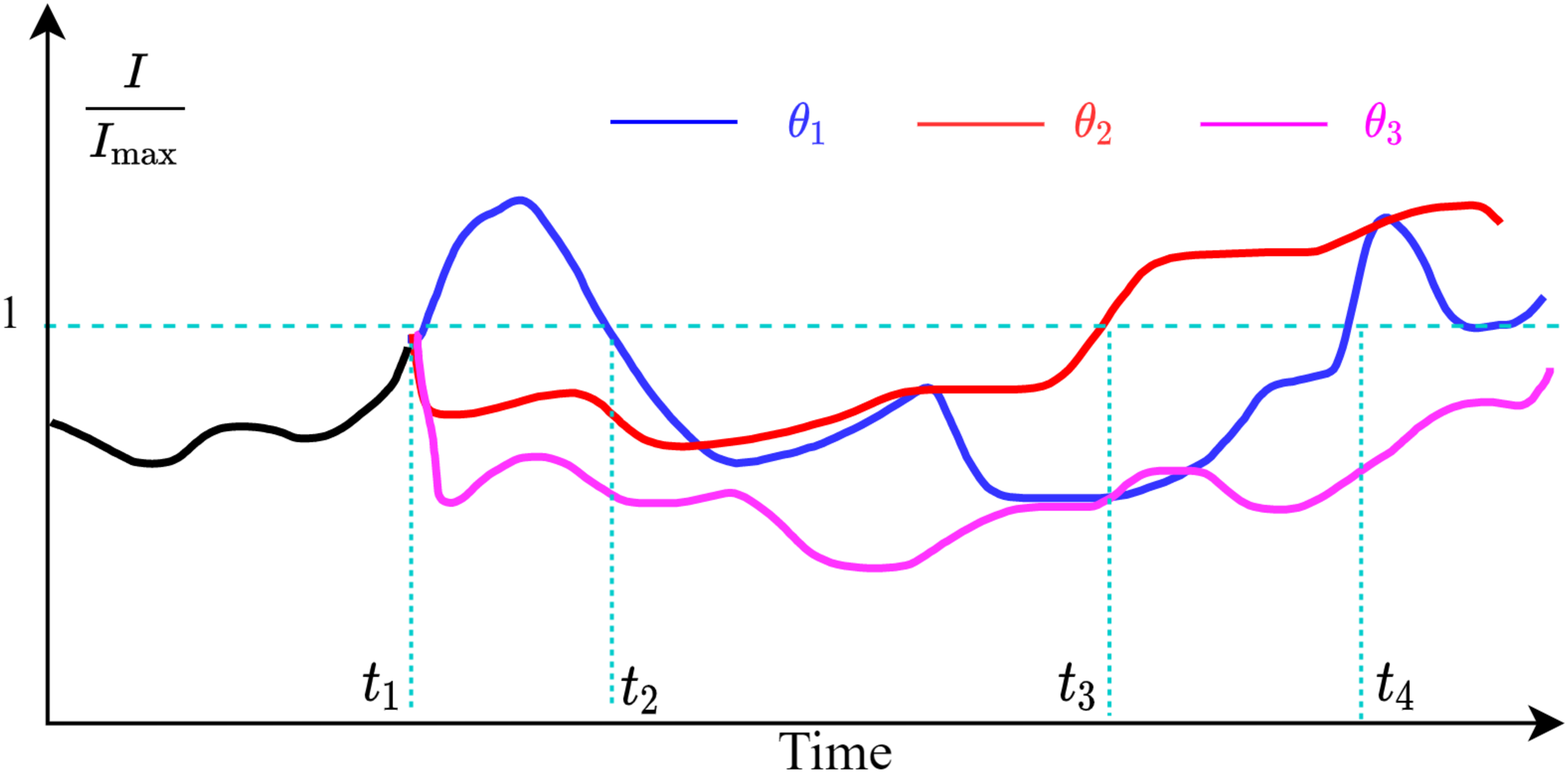}} 
    \caption{ \textcolor{black}{ Conceptual plot of the normalized line current versus time for three parameter values in a single scenario.}}
    \label{fig:concept_plot_episode}
    \vspace{-0.5cm}
\end{figure}

The relaxed line limit enforcement allows the operation of the grid for more time steps in an episode and generates more samples for training the agent at lower levels. Consider the situation shown in Fig \ref{fig:concept_plot_episode} displaying the normalized line current in a line for three-parameter values. An agent with any of the three parameters will see the entire scenario in level-1. The training samples for level-1 include samples where the agent's actions led to unfavorable/risky states. However, in level-2 an agent with the parameter $\theta_1$ will cause the environment to terminate 15 time-steps after $t_1$. A similar case occurs for $\theta_2$ at $t_3$. Hence, there are lesser samples from unfavorable/risky states with level-2 and level-3 enforcement. 

\begin{figure}[t]
    \centering
    \centerline{\includegraphics[width=\linewidth,trim={0.0cm 0.2cm 0.0cm 0.4cm},clip]{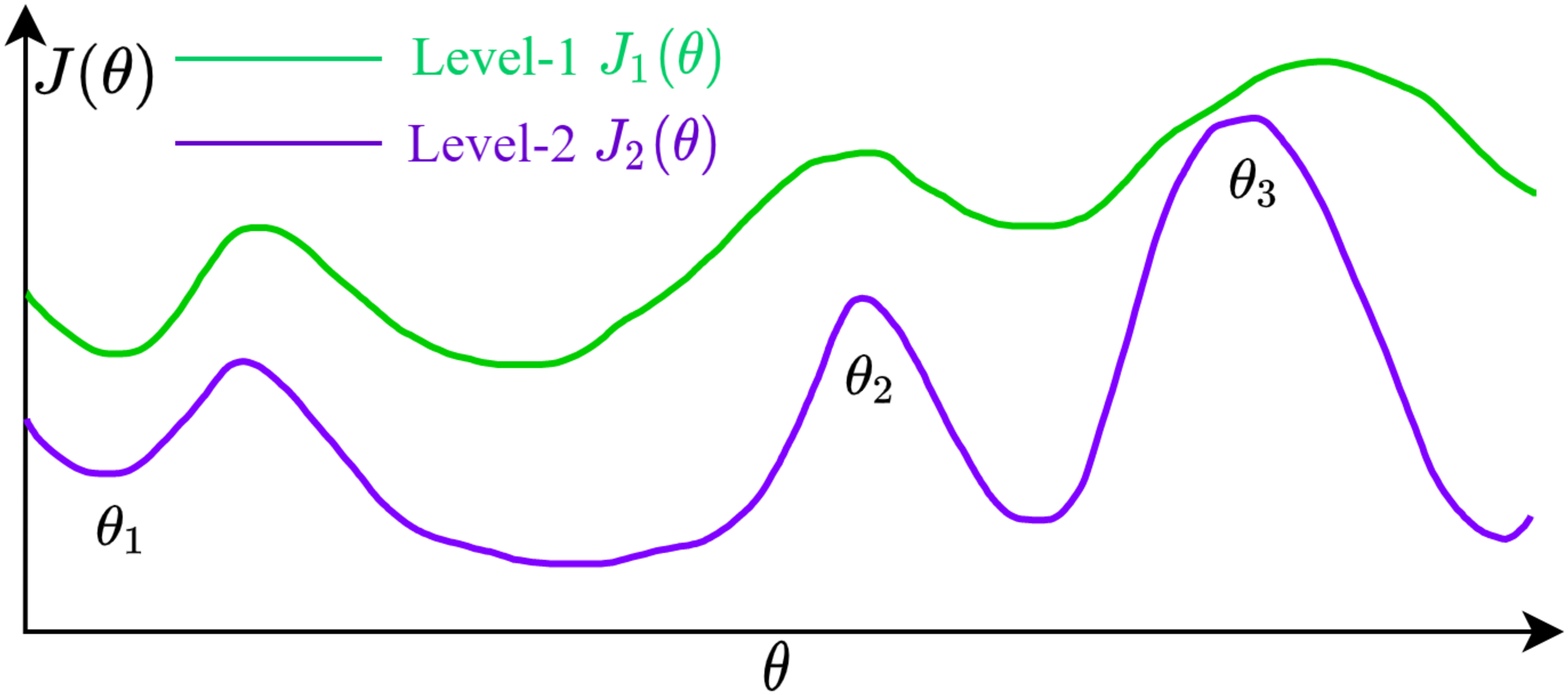}} 
    \caption{ \textcolor{black}{ Conceptual behavior of the objective function $J(\theta)$ versus varying parameters $\theta$ for different curriculum levels.}}
    \label{fig:concept_plot_obj_func}
    \vspace{-0.5cm}
\end{figure}


\subsection{Proposition 2: Agent learning for lower levels is easier than higher levels as the function $J(\theta)$ is smoother}

Effective local maxima hamper the agent learning in the objective function that prevents gradient-based methods from escaping. Curriculum learning smooths out the objective function and makes it easier for the optimization approaches to escape the local maxima. Consider Fig. \ref{fig:concept_plot_obj_func} which plots the conceptual objective function for two levels assuming one dimensional $\theta$. $J_2(\theta)$ has sharper peaks and troughs due to the fact that the enforcement of line currents is strict. Based on proposition 1, the relaxation of constraint enforcement in level-1 leads to a larger region of the parameter space where the agent performs well. The performance enhancement is highest for parameters that performed poorly in level-2 as they have the most room to improve ($\theta_1$ \& $\theta_2$ in this case). The performance of $\theta_3$ does not really improve as it is very successful in level-2. Hence, the overall effect on $J(\theta)$ is to reduce the variation between the peaks and troughs. Thus, the behavior of the eq. (\ref{obj_f}) is smoother for level-1(level-2) than level-2(level-3) over a larger parameter set, easing the agent learning in lower levels compared to higher levels.

\subsection{Proposition 3: Agents trained on lower levels perform well on higher levels as well}
The region in the parameter space which maximizes the objective $J_{1}(\theta)$ is close to the optimal value for the objective for $J_{2}(\theta)$. This is due to the design of the reward function, which penalizes overloads. An agent trained for sufficient episodes in level-1 will prevent large line overloads and avoid actions that island a part of the grid. Thus, an agent trained on level-1 for a level of success is likely to perform well on level-2 as well. Using the agent trained on level-1 as an initial agent for level-2 has the advantage of transferring learning from a simpler level as it does not need to relearn some action sequences. This property leads to efficient learning. The same logic holds while using an agent from level-2 to level-3. Hence, learning the agent sequentially from level-1 to level-3 ensures that the knowledge learned by the agent for 'easier' levels is retained, and the agents need fewer scenarios to satisfy the 'stricter' constraints of a 'harder' level. 

Hence, gradually increasing the level during the learning will lead to accelerated learning of the A3C agent \textcolor{black}{because of the above three prepositions as follows:} (1) it can observe and learn from more samples in lower levels and (2) gradient steps are more likely to skip the local maxima in the objective function in the relaxed levels (3) effective transfer learning occurs due to the design of the reward function. \textbf{The curriculum design is based on preventing cascading in any network. So, the curriculum described here can also be applied to other networks where the network flow is a key constraint. For example, computer networks where the routers can stop responding due to excessive traffic or in natural gas pipelines where the flow is constrained due to physical limits. Cascading occurs in these networks due to link/line overloading. Training RL agents in these domains to optimize network flow can benefit from this curriculum with minimal modifications.}

	\section{Simulation Results}
	\label{sec:Simulations}
In this section, results on the IEEE 14 bus system are presented. 
The load \& generation scenarios are taken the pypownet package \cite{lerousseau2021design} as a part of the Learn to Run the Power Network (L2RPN) 2019 challenge \cite{marot2020learning}. 

\subsection{Agent Training and  Evaluation Setup}
\label{subsec:train_and_eval_setup}
Deep neural networks represent both the actor and critic with two hidden layers of sizes 200 and 50. The first layer of the neural network is shared between the actor and critic leading to joint training of the A3C agent. The learning rate for the actor is 0.0005, and the learning rate for the critic is 0.001. A discount factor ($\gamma$) equal to 0.95 is used to calculate the time discounted rewards for the training.
A total of 50 {unique} training scenarios are selected from the dataset, and 50 threads are used in parallel during the A3C training procedure. Each unique scenario is made up of 2000 time steps of 5 minutes each that corresponds to 1 week of operation. \textcolor{black}{An agent that continuously operates the grid for all time steps in a scenario is categorized as a successful agent for that scenario.}

The following agents are used to verify the utility of reinforcement learning and curriculum learning to address the topology problem. There is no training in the forecasted power flow-based agent, as it is a brute-force approach, while the A3C agents are trained for 30,000 episodes on the 50 unique scenarios. 
\begin{itemize}
\item Forecasted power flow (FPF) based agent: This is the non-machine learning approach in which the forecasted injections at the next time step are used to identify the best action at a given time step. This approach is a 'greedy' approach as it is based only on a single-step forecast. It cannot account for how an action would change the line currents further into the future. 
    \item Baseline A3C (BA3C) Agent: This agent is trained on level-3 enforcement, the hardest level, using the modified reward with action/state-space reduction and state normalization throughout the training process.
    \item Curriculum A3C (CA3C) Agent: This agent is trained using the curriculum strategy presented in Section \ref{sec:curr_training_strategy} along with the modified reward with action/state-space reduction and state normalization. The transition between the levels occurs when the agent can continuously operate the grid for $>$1000 time-steps on at least 25 scenarios. 
\end{itemize}


\subsection{Training of the BA3C and CA3C Agents}
\begin{figure}
    \centering
    \centerline{\includegraphics[width=\linewidth]{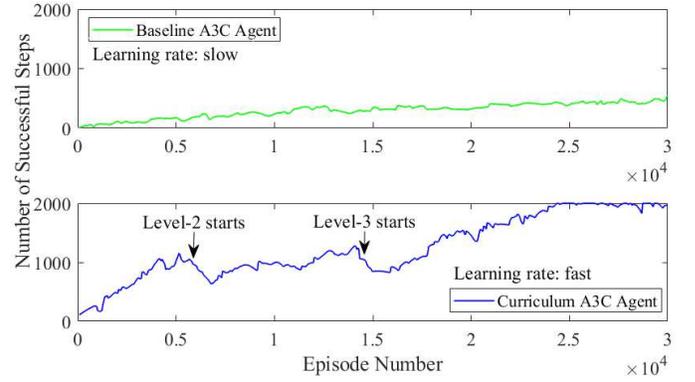}} 
    \caption{ \textcolor{black}{Plot of the number of successful steps versus the episodes during training of BA3C and CA3C agents.}}
    \label{fig:training_scores}
    \vspace{-0.2in}
\end{figure}

The agents are implemented in Keras and are trained using TensorFlow for 30,000 episodes {and the code is available on GitHub \cite{pypownet_a3c}}. The number of successful time steps at each training episode for the two agents is shown in Fig.~ \ref{fig:training_scores}. The median {of successful time steps for each of the 30,000 episodes} over a window of 15 different scenarios/weeks is plotted in Fig.~ \ref{fig:training_scores} to smooth out the large variation among the episodes. The enforcement level of CA3C is initially level-1. Based on the agent's performance, the enforcement level is increased to level-2 at episode 6000 and increased to level-3 at episode 14000. The agent at these episodes is saved for the further analysis presented in subsection-D.

It can be seen from the plot in Fig.~ \ref{fig:training_scores} that the learning is comparatively slow for the BA3C agent. The number of successful steps of the BA3C agent in the training phase at the end of 30,000 episodes is around 500 steps. For the CA3C agent, there is a much faster training rate as the number of successful steps increases quickly. This is because the enforcement level is low (level-1). As soon as the level is increased after 6k episodes, there is a drop in the number of successful steps. After a few more episodes, the learning algorithm will update the network parameters appropriately and improve them till the next level is enforced at 14k episodes. The same temporary drop in performance can be seen after 14k episodes. It was observed that the variance of the rewards observed during BA3C training is higher compared to CA3C training. This is due to the 'rough landscape' of the objective function $J(\theta)$ for level-3 enforcement.   
\subsection{Evaluation of Various Agents on Test Scenarios}
\begin{figure}
    \centering
    \centerline{\includegraphics[width=\linewidth]{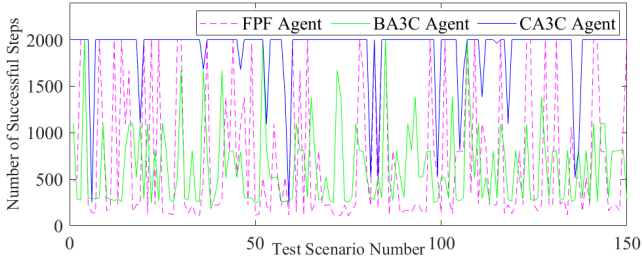}} 
    \caption{The number of successful steps for the various agents in each test scenario with level-3 enforcement. Each scenario is a unique one week worth of operating conditions.}
    \label{fig:testing_steps}
    \vspace{-0.5cm}
\end{figure}

In this section, the performance results of the various agents are presented and analyzed. The three (FPF, BA3C, and CA3C) agents are evaluated on 150 test scenarios enforced at the hardest level, with each scenario lasting 2000 time steps. The agents are used to identify topology actions only during the time-steps when the current flow in at least one line exceeds 80\% of its limit. The agents are scored on each scenario based on the number of continuous successful time steps before the scenario terminates due to islanding or system divergence. This information is plotted in Fig.~\ref{fig:testing_steps}. 

\begin{figure}
    \centering
    \centerline{\includegraphics[width=\linewidth,trim={0.0cm 0.2cm 0.0cm 0.2cm},clip]{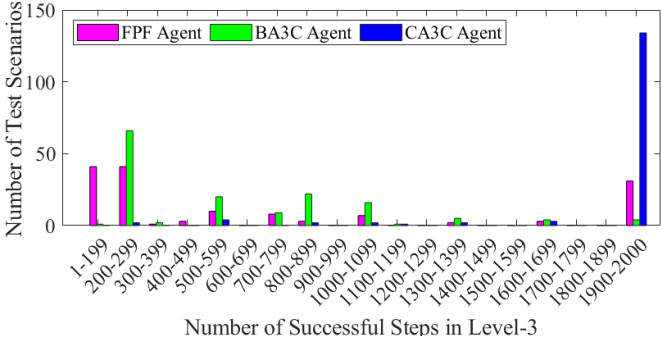}} 
    \caption{Histograms of the successful steps for the various agents for 150 test scenarios with level-3 enforcement.}
    \label{fig:histogram_comparison_hard}
    \vspace{-0.5cm}
\end{figure}

The performance of the BA3C agent is poor as only a few of the scenarios successfully reached the end. Most of the scenarios with the BA3C agent terminated within 500 time-steps. In contrast, the performance of the CA3C agent is much better as most (120 out of 150) of the scenarios successfully reached the end. The behavior of the three agents is summarised in Fig.~\ref{fig:histogram_comparison_hard} which plots the histogram of the number of successful time steps for each test scenario partitioned into bins of 200-time steps. The superior performance of the CA3C agent can be clearly seen from this histogram. These results demonstrate that (i) reinforcement learning agents can perform better than a single look-ahead non-ML-based approach on systems with complex constraints on the actions (ii) The CA3C agent performs significantly better than the BA3C agent trained without a curriculum.


\subsection{Illustrating Accelerated Learning Due to Curriculum}
\begin{figure*}[ht]
    \centering
    \centerline{\includegraphics[width=0.875\linewidth]{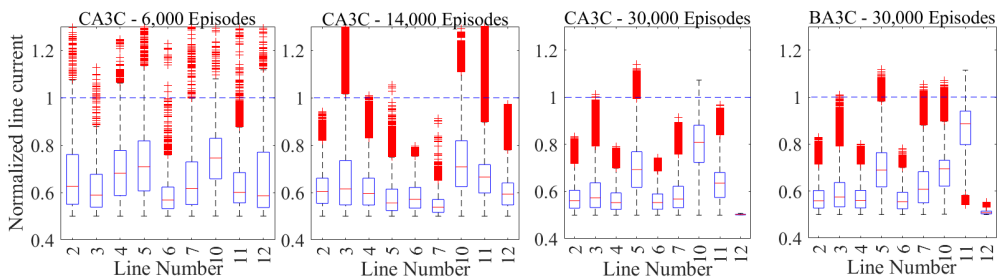}} 
    \caption{Comparison between the boxplots of the line currents in each line due to the actions of the various A3C agents for all the time steps in the 150 test scenarios with over-current cascading disabled.}
    \label{fig:boxplot_rho}
    \vspace{-0.1in}
\end{figure*}

To demonstrate why the {proposed approach-based} curriculum training leads to a better agent, we analyze the behavior of the A3C agents on the test scenarios with level-1 enforcement. In addition, we also evaluated the behavior of the CA3C agents stored when the enforcement levels were raised. CA3C-6k is the agent when the curriculum transitions from level-1 to level-2 at episode 6000, and CA3C-14k is the agent when the curriculum transitions from level-2 to level-3 at episode 14000. We recorded the normalized line currents for all the lines for each agent and scenario, leading to a large database. A box-whisker plot is used to present the statistics of the dataset visually and is shown in Fig.~\ref{fig:boxplot_rho}. The red line within each box is the median current flow for all test scenarios for a particular A3C agent. The top and bottom boundaries of the box correspond to the inter-quartile range of the current for each line. The top and bottom whiskers correspond to the estimated maximum and minimum values of the current for each line without the outliers. The outliers of the current for each line are represented as red crosses. 

By observing the box plots in Fig.~\ref{fig:boxplot_rho}, it can be seen that the CA3C-6k agent has many samples with overload. This is expected as the agent has not yet learned fully to reduce the line currents. Instead, the agent has prioritized the identification of action sequences that would lead to termination due to islanding without any cascading. 
In the box plot of the CA3C-14k agent, we can observe that the agent has learned to reduce the line current below the maximum value for many of the lines. This is exactly the intention behind increasing the value of $\alpha$ and adding a time delay to the overload disconnection. 
Finally, after 30,000 episodes, the CA3C agent reduces the overload to just 2 lines. In contrast, the BA3C agent after 30,000 episodes has overloads in 4 lines, making it more susceptible for cascading. 

\begin{figure*}[ht]
    \centering
    \centerline{\includegraphics[width=0.875\linewidth]{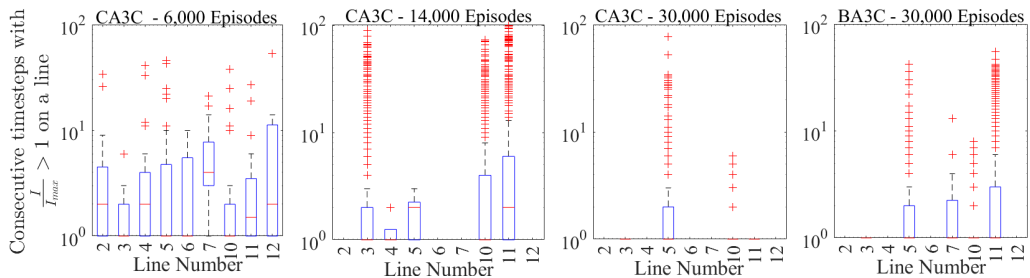}} 
    \caption{Comparison between the boxplots of the number of consecutive time-steps for over-currents in each line due to the actions of the various A3C agents for all the time steps in the 150 test scenarios with over-current cascading disabled.}
    \label{fig:boxplot_cons_timestep}
    \vspace{-0.2in}
\end{figure*}

However, this analysis does not fully explain the poor performance of the BA3C agent, as most of the lines have avoided overloading. This is because the number of time steps that a line is continuously overloaded ($t_{p,over}$) is the actual reason for line disconnection, and this is not the same as the total number of time steps that a line is overloaded. For example, an agent that can immediately rectify a line overload in one step will have a $t_{p,over}$ equal to 1. Thus, a successful agent has smaller values ($t_{p,over}$) for all lines. The data set is analyzed, and $t_{p,over}$ is calculated. The resulting $t_{p,over}$ are plotted in Fig.~\ref{fig:boxplot_cons_timestep}. 

The box plots in Fig.~\ref{fig:boxplot_cons_timestep} demonstrate that the time spent by each line continuously in overload reduces as the CA3C agent learns. Initially, many lines have high $t_{p,over}$. As the learning progresses, the value of $t_{p,over}$ reduces. At the end of 30,000 episodes, the CA3C agent can limit $t_{p,over}$ to 1 for all lines except line-5. These values of $t_{p,over}$ are low enough that most overloads do not cause line disconnections, limiting the impact of most of the overloads. In contrast, the maximum inter-quartile value of $t_{p,over}$ for the final BA3C agent is equal to 3 for line 5, 4 for line 7, 1 for line 10, and 8 for line 11. These values are much larger than the HOT, and thus they will lead to cascades in most test cases. This is exactly what we observe in Fig.~\ref{fig:testing_steps} for the BA3C agent. 

Hence, the CA3C agent can minimize the overloading occurrences and also reduce the continuous-time in the overloaded state, thus leading to improved performance compared to the BA3C agent. \textbf{The efficient learning of the CA3C agent is verified on the IEEE 14 bus system by systematically analyzing the cause of the failure of the partially trained CA3C agents and the BA3C agent. The statistical analysis of the line currents and consecutive duration of the line overloads is used to justify the gradual improvement in the performance of the CA3C agent as training proceeds. This is the third contribution of the paper.}  

\subsection{Agent Performance in L2RPN-2019 Competition}

\textbf{The A3C agent trained using the curriculum presented in this paper placed 2nd in the L2RPN-2019 challenge by RTE.} RTE tested the trained CA3C agent on hidden scenarios of varying length, and the agent was successful in all the cases. RTE's analysis from \cite{marot2020learning} for the trained CA3C agent mentions that the agent is quite stable due to its small action
space but has the ability to go back and forth, illustrating the impact of using the physics of the system in the designing of the action space. \textbf{The authors have open-sourced the code to train the A3C agent with the physics-based curriculum approach for controlling grid topology on GitHub from \cite{pypownet_a3c}. This is the fourth and final contribution of the paper.} The novel contributions of the paper (state-space reduction, action space reduction, modified reward, and novel curriculum training methodology) and the corresponding code has been already used by another team to win the L2RPN-2020 challenge on a larger system \cite{grid2op_third_place}, demonstrating the applicability of the proposed approach for larger systems.
%
	\section{Conclusion}
	\label{sec:conclusion}
	This paper describes how domain knowledge of power system operators can be integrated into reinforcement learning frameworks to effectively learn agents that control the grid to prevent cascading through grid reconfiguration. The non-linear and combinatorial nature of the grid reconfiguration problem means that no existing optimal power flow solver can yet tackle this problem. We have developed an actor-critic-based agent that has successfully operated the grid under various test scenarios. The key to training this agent is to incorporate the knowledge of power system operation into various aspects of the reinforcement learning framework. By analyzing the grid topology and grid operation, the action space and the state space dimensions are significantly reduced; and a reward function is designed to provide gradients even when the grid has overloaded. Furthermore, an effective curriculum-based approach incorporated into the training procedure through environment modifications enables the agent's accelerated learning. The learning procedure is stabilized and made robust to the natural variability in grid operations by employing a parallel training procedure that trains on multiple scenarios of the power grid at the same time. This reduces the sampling bias that is likely to seep through when training using a sequential training method. Without these enhancements to the training procedure, the RL agent failed for most test scenarios, illustrating the importance of properly integrating domain knowledge of the physical system for RL learning for a real-world system. The developed code is available online and open-sourced for public use.
	\bibliographystyle{IEEEtran}
	\bibliography{main}

\end{document}